# ENVIRONMENT-AWARE RECONFIGURABLE NOISE SUPPRESSION

*Jun Yang (IEEE Senior Member) and Joshua Bingham*

Facebook AR/VR Division, 1 Hacker Way, Menlo Park, CA 94025, USA

**ABSTRACT**

The paper proposes an efficient, robust, and reconfigurable technique to suppress various types of noises for any sampling rate. The theoretical analyses, subjective and objective test results show that the proposed noise suppression (NS) solution significantly enhances the speech transmission index (STI), speech intelligibility (SI), signal-to-noise ratio (SNR), and subjective listening experience. The STI and SI consists of 5 levels, i.e., bad, poor, fair, good, and excellent. The most common noisy condition is of SNR ranging from -5 to 8 dB. For the input SNR between -5 and 2.5 dB, the proposed NS improves the STI and SI from "fair" to "good". For the input SNR between 2.5 to 8 dB, the STI and SI are improved from "good" to "excellent" by the proposed NS. The proposed NS can be adopted in both capture and playback paths for voice over internet protocol, voice-trigger, and automatic speech recognition applications.

## 1. INTRODUCTION

Audio devices have been playing an important role in our daily life. However, our environment is full of various types of noises. All these noises can significantly degrade the speech quality, as well as the processing performance related to acoustic echo cancellation (AEC), voice-trigger, automatic speech recognition (ASR), and voice over internet protocol (VoIP) [1 – 10].

The noise could be either stationary, non-stationary, or their complex combinations. However, traditional background NS approaches assume that noise is stationary and hence cannot suppress all other types of noises in an efficient way. Also, some existing NS algorithms need the voice activity detection (VAD) and estimate the noise during speech pause. Hence, they greatly depend on the efficiency and accuracy of the VAD. In fact, VAD is a very difficult [11, 12] and still ultimately unsolved problem for realistic situations where the noise level keeps varying with time.

Challenging conditions for NS also include but not limited to low SNR, fast-varying probability distributions, nonlinear combinations of different types of noises, etc. In facing these adverse conditions, the existing NS approaches can improve SNR but will unfortunately generate speech distortions, unnatural sounding or fluctuating residual noises although they have improved SNR. More importantly, these existing algorithms fail to provide the required reduction accuracies when the desired voice and the noise simultaneously present, which is the case in many practical applications.

Therefore, the computationally cost-effective and robust techniques that can handle more accurately and more efficiently various types of noises are highly desirable so as to provide the needed performance improvement for applications related to voice-trigger, ASR, and VoIP. It is the goal of this paper to provide such a desired NS solution as described in the next sections.

The rest of this paper is organized as follows. Section 2 presents the overview of traditional NS schemes and their shortcomings. The focus of Section 3 is on describing all the processing steps of the proposed NS algorithm. The simulation results mainly presented in Section 4 will show that the proposed NS solution can significantly improve STI, SI, SNR, and subjective listening experience so that VoIP, barge-in, and ASR performance can be improved correspondingly. As the last section of this paper, Section 5 will make some conclusions and further discussions.

## 2. TRADITIONAL NOISE SUPPRESSION SCHEMES

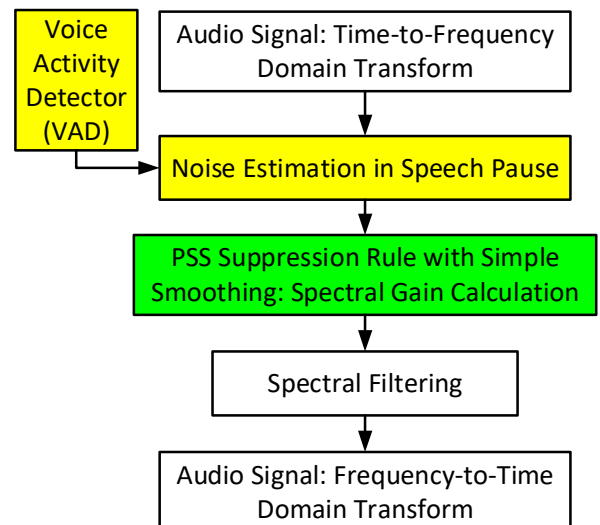

Figure 1: Block Diagram of Traditional NS Schemes.

Figure 1 shows the block diagram of the most-commonly used traditional NS schemes. One key step is the VAD based "Noise Estimation in Speech Pause" which means that the noise spectrum is estimated when speech is absent.

Apparently, these NS schemes highly rely on the VAD performance and cannot accurately estimate the noise spectrum when SNR is low.

The suppression rule is defined by the following equation.

$$G(m, k) = [1 - (\frac{Y(m,k)}{|X(m,k)|^2})^{\frac{\alpha}{2}}]^{\beta} \quad (1)$$

where $X(m, k)$ and $Y(m, k)$ are input spectrum and the estimated noise power spectrum at the *k-th* bin (or *k-th* band in subband domain) and in the *m-th* frame. The $\alpha$ and $\beta$ are two constants which determine the type of traditional parametric spectral subtraction (PSS) methods. The Table 1 gives their definitions. The values of the obtained spectral gain is required to satisfy $0.0 < G(m, k) \leq 1.0$

Table 1: Traditional PSS Noise Suppression Schemes

| Method | (α, β) |
|---|---|
| Power Spectral Subtraction | (2, 1/2) |
| Magnitude Spectral Subtraction | (1, 1) |
| Short-time Wiener Filtering | (2, 1) |

As mentioned in the first section, these traditional NS schemes can result in the artifact called "musical noise" which mainly consists of a succession of randomly distributed spectral peaks and sounds like a sequence of pure tones. To remove this artifact and deliver a better noise suppression has been a major goal in the related research and development fields. To achieve this goal, this paper presents a new scheme with mainly proposing an advanced suppression rule and adaptive smoothing method as detailedly described in Section 3.

### 3. THE PROPOSED NOISE SUPPRESSION SCHEME

Figure 2 illustrates the schematic diagram and data flow of the proposed environment-aware reconfigurable NS algorithm. The details of its major processing steps are given as follows.

The input audio signal is firstly processed by a reconfigurable high-pass filter (HPF) to reduce low frequency noises. The cutoff frequency of HPF can be configured to be ~80 Hz for wideband VoIP, voice-trigger, and ASR applications, and configured to be ~150 Hz for narrowband VoIP application.

With 50% overlap, a frame of the filtered signal with frame size *N* samples is constructed by using *N/2* samples of the current frame and *N/2* samples of the previous frame. For the case that overlap is not 50%, zero-pad approach is used to construct a frame of size *N* samples.

After windowing and N-point FFT processing on the time domain signal, the signal is then transferred into frequency domain from the original time domain. The power spectral density (PSD) is further computed for *N/2* frequency bins. It should be noted that the frequency bin 0 is DC component (i.e., the mean amplitude of the waveform) and could be set as zero level, frequency bin *N/2* is Nyquist component (i.e., $f_s/2$ component. The $f_s$ is a sampling rate).

According to different application environments and settings, the window function can be reconfigurable from Hanning window, Hamming window, the raised-cosine window, and so on.

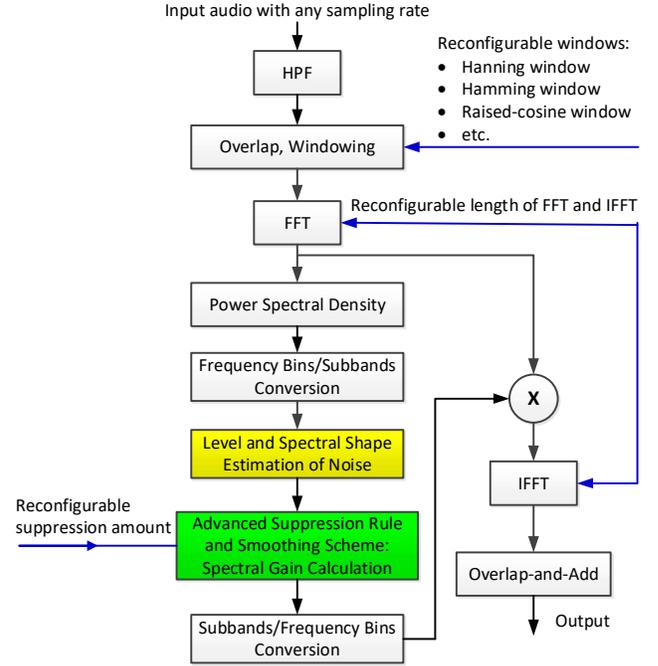

Figure 2: The proposed Environment-Aware Reconfigurable NS algorithm.

To reduce computational complexity and memory usage, the PSD of *N/2* bins are partitioned into *M* bands (e.g., *M* = 36) on the basis of the psychoacoustic critical bands. The root-mean-square (RMS) level is calculated for each band. For comparison, traditional power spectrum and amplitude spectrum are typically used in the corresponding power subtraction and magnitude subtraction, respectively.

The noise floor estimation and dynamic attenuation gain calculation will be conducted on the basis of the obtained subband RMS spectrum.

The proposed block of "Level and Spectral Shape Estimation of Noise" in Figure 2 is implemented by searching the statistic minimum RMS level of each frequency band over a sliding time-window so that background noise (both level and spectral shape) is continuously updated and controlled, which is a key component to enable suppressing the nonstationary noise. The length (e.g., 2 seconds) of the sliding time-window determines the convergent time.

To generate natural sounding output signal and avoid distortion, we propose to smooth the obtained noise RMS spectrum over time as follows.

$$W(m, k) = W(m-1, k) + \alpha * (W'(m, k) - W(m-1, k)) \quad (2)$$

where $W'(m, k)$ is the estimated noise RMS spectrum at the *k-th* band (*k = 0, 1, ..., M-1*) and in the *m-th* frame, $W(m, k)$ is the smoothed noise RMS spectrum, parameter $\alpha$ is a smoothing factor and ranges from 0.0 to 1.0.

For the purpose of reducing computational complexity, a method to calculate the spectral gain in subband domain is developed in this paper. More importantly, this developed spectral gain calculation method can also reduce the speech distortion and residual noise distortion. The following are the details of the proposed calculation of the spectral gain in subband-domain.

$$G'(m, k) = \sqrt{1 - \mu \frac{|W(m,k)|^2}{|Z(m,k)|^2}} \quad (3)$$

where $|Z(m, k)|$ is the input subband RMS spectrum at the $k$-th band and in the $m$-th frame, the value of parameter $\mu$ is between 0.0 and 1.3. Noise is underestimated when $\mu$ is between 0.0 and 1.0 and is overestimated when $\mu$ is between 1.0 and 1.3. The larger is the value of parameter $\mu$, the more noise is suppressed. The values of spectral gain is required to satisfy $maxSuppression \leq G'(m, k) \leq 1.0$. The value of parameter $maxSuppression$ is reconfigurable according to the different applications. For example, the parameter $maxSuppression$ can be configurable to be 0.5 and 0.1259 for voice-trigger and ASR application (i.e., maximum 6 dB noise suppression) and VoIP application (i.e., maximum 18 dB noise suppression), respectively.

The obtained spectral gain could vary dramatically from frame to frame, which will generate unnatural sound and distortion (such as "musical noise"). To generate natural sounding output signal and avoid such distortion, what we propose is to modify the obtained spectral gain as follows.

$G(m, k) = G(m-1, k) + β(m, k) * (G'(m, k) - G(m-1, k))$
        for $0 \leq k \leq M-1$             (4)

where $β(m, k)$ is a time-frequency dependent smoothing factor satisfying $0.0 \leq β(m, k) \leq 1.0$. The variable $β(m, k)$ can be reconfigurable such that the larger is the $G'(m, k)$, the larger value is set for the $β(m, k)$.

After converting the obtained $M$ band gains into the $N/2$ spectral bin gains $P(m, i)$ ($i = 1, 2, ..., N/2$), we dynamically apply the $N/2$ bin gains $P(m, i)$ to the original signal spectrum by using the original phase response.

Further performing IFFT and overlap-and-add processing, a frame signal in time domain is finally constructed with noise suppressed to the desired degree.

To demonstrate the effectiveness of the proposed solution and the accuracy of the related analyses, two types of acoustical evaluations and related settings will be reported in the next section.

## 4. EVALUATION RESULTS

This section presents evaluation results of the proposed environment-aware reconfigurable NS algorithm in terms of STI, SI, SNR improvement, and subjective listening tests.

### 4.1. STI and SI performance

Figures 3 and 4 show the STI scores and SI scorers, respectively, before and after our NS processing. The STI and SI scores are obtained by using the modified rhyme tests (MRT) (word perception) according to the standard IEC 60268-16 [13, 14]. The black curves in Figures 3 and 4 are obtained from the original noisy speech data, the blue curves are obtained from our NS processing with 6 dB maximum suppression setting, the red curves are obtained from our NS processing with 12 dB maximum suppression setting.

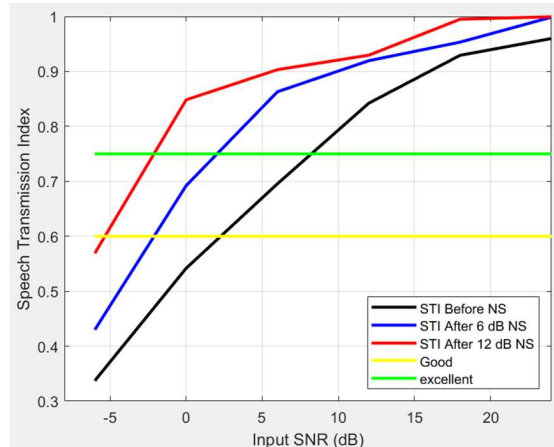

Figure 3: STI Scores Before and After Our NS Processing.

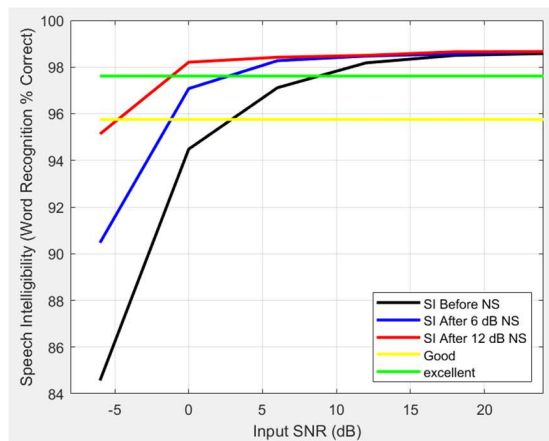

Figure 4: SI Scores Before and After Our NS Processing.

If STI scores are above its green line, then the STI scores are considered as "excellent". If scores are between green line and yellow line, then STI scores are considered as "good". Otherwise, STI scores are not acceptable. This is the case for SI scores as well.

It can be seen from Figures 3 and 4 that our NS scheme with both 6 dB suppression setting (blue curves) and with 12 dB suppression setting (red curves) has significantly improved the STI and SI for the case of the input SNR larger than -6 dB.

More importantly, the most common noisy condition in real life is of SNR ranging from -5 to 8 dB. For the input SNR between -5 and 2.5 dB, the proposed solution with 12 dB NS setting improves the STI and SI from "fair" to "good". For the input SNR between 2.5 to 8 dB, the STI and SI are improved from "good" to "excellent" by the proposed 12 dB NS settings.

## 4.2. SNR improvement and subjective listening tests

We have run 72 test cases by covering the following test conditions:
(1) two input SNRs: 6 dB and 12 dB,
(2) two speech-levels at mouth reference point: 89 and 95 dBC,
(3) two distance settings between device-under-test (DUT) and head-and-torso-simulator (HATS): 1 meter and 4 meters,
(4) 9 types of noises: air condition noise, café noise, fan noise, living-room noise, office noise, pink noise, Pub noise, rain noise, and rock musical noise.

Figures 5(A) and 6(A) show the input waveform (top) and spectrogram (bottom) of speech in air condition noise and living-room noise, respectively. Figures 5(B) and 6(B) show the corresponding output waveform (top) and spectrogram (bottom) obtained by our NS processing. It should be noted that the magnitude scales of waveforms in Figures 5(B) and 6(B) are the same as those in their corresponding input waveforms. Obviously, our NS algorithm significantly suppresses these noises and enhances the SNR performance.

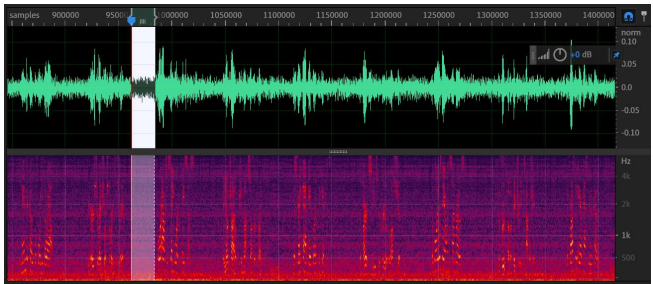
Figure 5(A): Waveform (top) and Spectrogram (bottom) before NS Processing, Speech in air condition noise.

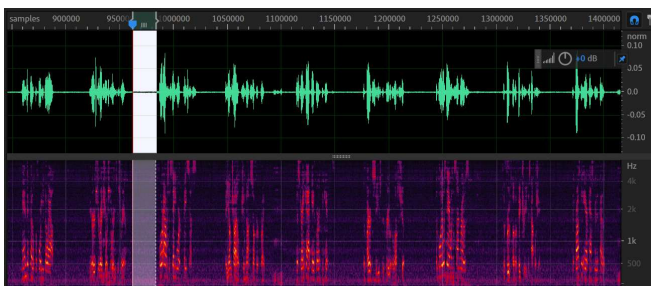
Figure 5(B): Waveform (top) and Spectrogram (bottom) after Our NS Processing.

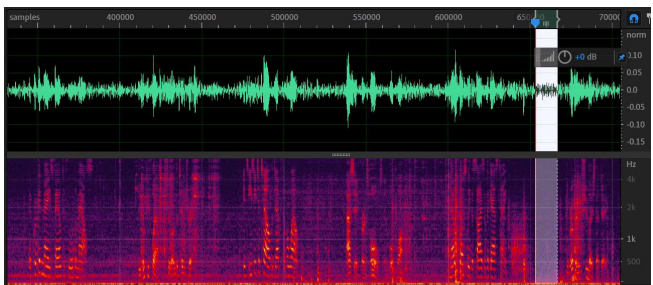
Figure 6(A): Waveform (top) and Spectrogram (bottom) before NS Processing, Speech in living-room noise.

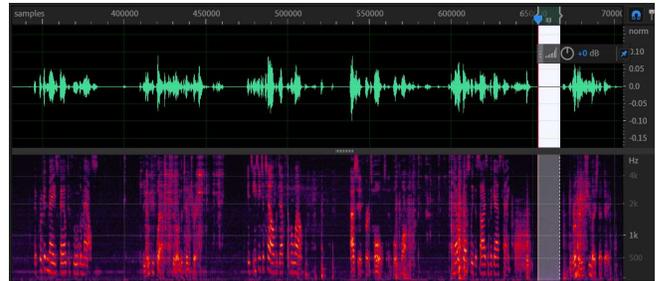
Figure 6(B): Waveform (top) and Spectrogram (bottom) after Our NS Processing.

It is worth emphasizing that:
(A). the air condition noise is reduced by ~24 dB after our NS processing according to Figures 5(A) and 5(B).
(B). the living-room noise is reduced by ~37 dB after our NS processing according to Figures 6(A) and 6(B).

In addition to the above objective tests, extensive subjective listening test results have shown that speech perception is significantly improved in noise. It has been also demonstrated from these testing that the processed speech and the residual noise are able to sound both high quality and natural way.

## 5. CONCLUSONS

A new environment-aware NS solution implemented via C reference code has been reported in this paper, which can improve all key performance indicators with including speech intelligibility, speech transmission index, signal-to-noise ratio, and subjective listening experience.

This proposed NS algorithm can also be used for multi-microphone system in which each microphone signal could have the independent parameter settings.

Furthermore, this proposed NS processing can support for any sampling rate and for any frame size. Without introducing artifacts (such as, musical noise) the proposed solution can be reconfigurable in such a way as to support various applications and products with different types of noises and parameters settings. The related parameters can be predefined and tuned by the related corresponding training datasets.

## 6. ACKNOWLEDGMENTS

We wish to express our sincere thanks to Jens Nilsson (Facebook Audio Technology Engineering team @ Portal) for providing noisy speech database. We also express our special thanks to Gongqiang Yu (Facebook Audio Technology Engineering team @ AR) for providing STI and SI evaluation tool, clean speech data, and noise data.